\documentclass{article}
\begin{document}

\begin{center}
{\LARGE On the Equilibrium Shape of an Ice Crystal}\vskip6pt

{\Large Kenneth G. Libbrecht}\vskip4pt

{\large Department of Physics, California Institute of Technology}\vskip-1pt

{\large Pasadena, California 91125}\vskip-1pt

\vskip18pt

\hrule\vskip1pt \hrule\vskip14pt
\end{center}

\textbf{Abstract}: We examine the shape of a an isolated, dislocation-free
ice crystal when it is in equilibrium with the vapor phase in an isothermal
closed environment, as a function of temperature. From our analysis we draw
the following conclusions: 1) The equilibrium shape has not yet been
definitively measured for ice crystals; 2) The surface energy anisotropy is
likely cusp-like near the facet angles, and the size of the cusps can be
estimated from crystal growth measurements; 3) The equilibrium shape is
likely nearly spherical with only small faceted regions; 4) The time needed
to reach equilibrium is likely prohibitively long, except under special
circumstances; and 5) Surface energy effects likely play a relatively
smaller role in ice crystal growth dynamics when compared to the role of
attachment kinetics.

\section{Introduction}

Studies of the equilibrium crystal shape (ECS) for a given material are
useful for understanding several static surface properties, such as surface
energy anisotropies and step energies, and can also yield important insights
into dynamical properties such as step-step interactions, surface diffusion,
and crystal growth behavior. For this reason there has been much interest in
ECS measurements and theory over the past several decades, and there is a
considerable literature on the subject (for example see the review articles 
\cite{bonzel, nagaev, degawa}).

For many simple metals that have been studies, the surface energy
anisotropies are fairly small, typically a few percent difference between
surfaces of maximum and minimum energy \cite{heyraud1, heyraud2, chatain1,
metzmann}. The typical ECS is nearly spherical with small flats at the
lower-surface-energy facet surfaces, corresponding to cusps in the surface
energy (i.e. the surface energy $\gamma (\hat{n})$ has a discontinuous first
derivative at each facet angle, where $\hat{n}$ is the surface normal unit
vector). The ECS can be derived from $\gamma (\hat{n})$ using the well-known
Wulff construction, showing that the size of the cusps relates to the size
of the corresponding facet surfaces in the ECS. Adsorbates may also
substantially change $\gamma (\hat{n})$ and the ECS \cite{chatain2}.

We consider here the ECS of an isolated, dislocation-free ice crystallite
when it is in equilibrium with the vapor phase in an isothermal closed
environment, as a function of temperature. The questions we wish to address
include:

1) What is known about the ice ECS and surface energy anisotropies?

2) What experiments have been done to date, and what have they revealed?

3) What are the prospects for new experimental studies?

4) What role does surface energy play in ice growth from water vapor?

\section{Lowest Order Theory}

For a vicinal surface tilted a small angle $\theta $ from the facet angle,
we can write the surface energy as $\gamma (\theta )\approx \gamma
_{facet}+\gamma _{steps}(\theta ),$ where $\gamma _{facet}$ is the surface
energy of a flat facet surface and $\gamma _{steps}(\theta )$ is the
additional surface energy that arises from molecular steps on the surface 
\cite{Bonzel}. To lowest order, neglecting step-step interactions, this
becomes%
\[
\gamma (\theta )\approx \gamma _{facet}+\frac{\beta }{a}\sin \theta 
\]%
where $\beta $ is the step energy per unit length and $a\approx 0.32$ nm is
the step height. Plugging this into the Wulff construction then yields the
half-angle of the facet surface (as seen from the center of the crystallite)
for the ECS \cite{Bonzel} 
\[
\theta _{HW,facet}\approx \frac{\beta }{a\gamma _{facet}}. 
\]

For the surface energy of a basal or prism facet we estimate $\gamma
_{facet} $ $\approx 0.08$ J/m$^{2}$ \cite{petrenko, hobbs, pruppacher}, and
measurements of crystal growth yield $\beta \approx 1\times 10^{-12}$ J/m
for temperatures between $-2$ and $-10$ C \cite{libbrechtreview, libbrecht},
giving $\theta _{HW,facet}\approx 0.04$ radians. If two basal facets and six
prism facets are present with similar $\theta _{HW,facet}$ values, then the
total fraction of the ECS surface area that is faceted is approximately $%
f\approx 8\pi \theta _{HW,facet}^{2}/4\pi ^{2}\approx 0.1$ percent. This
fraction increases at lower temperatures, as $\beta $ is a strong function
of temperature, and at $-40$ C we find $f\approx 2$ percent. The biggest
uncertainty in this calculation is our knowledge of the step energy $\beta ,$
which has been derived indirectly, from 2D nucleation dynamics in crystal
growth observations.

Overall the surface energy $\gamma (\hat{n})$ of the ice-vapor interface is
not well known, either on the facet surfaces or on unfaceted surfaces.
Determining the absolute value of $\gamma (\hat{n})$ is difficult, and
experiments to date yield have yielded results with substantial
uncertainties. Calculations are also quite uncertain because the detailed
surface structure of ice is rather poorly understood, especially surface
melting and other surface restructuring, and this strongly affects estimates
of $\gamma (\hat{n})$. The anisotropy of $\gamma (\hat{n})$ could be
obtained from observations of the ECS, but experiments to date are not
definitive (see below). Nevertheless, in my opinion the measurements do
indicate with some confidence that $\beta \ll a\gamma _{facet},$ and from
this it appears likely that the ice/vapor ECS at temperatures $-40<T<0$ C
will exhibit only small faceted regions. It also appears that $\gamma
_{basal}\approx \gamma _{prism}$ (with an uncertainly of perhaps 20 percent,
although it could be greater \cite{petrenko, hobbs, pruppacher}), which
means that the ECS is likely nearly spherical.

We note that several textbooks suggest that the ECS for ice/vapor is fully
faceted \cite{pruppacher, hobbs, fletcher}. This appears to be a
misapplication of the Wulff construction, as the authors of these texts
mainly considered the surface energy of the facets, even though the angular
dependence of $\gamma (\hat{n})$ in its entirety is needed to calculate the
ECS. In addition, the experiments cited in these texts as evidence for a
faceted ECS were likely observing growth forms. The description of the ECS
is similar in these three references, so it may have been propagated from
one to the next over time.

We also note in passing that Landau has suggested that the equilibrium
shapes of all crystals should be completely faceted at $T=0$ \cite{landau,
Ramanujan}, a statement that is occasionally mentioned in the literature.
However it appears that this theoretical result is unphysical for several
reasons:\ 1) The time needed to reach equilibrium near $T=0$ is so long as
to be unrealizable in practice, and 2) the Landau model indicates that the
surface energy $\gamma (\hat{n})$ has a cusp for all rational Miller
indices, which would mean that $\gamma (\hat{n})$ is continuous everywhere
but does not possess a well-defined derivative anywhere \cite{Ramanujan}.
Thus the Landau theory seems to have little application to real experimental
systems.

\section{Equilibration Times}

An important consideration for experimental studies of the ECS is the time
required to reach equilibrium. For the case of ice, the equilibration
process is somewhat similar to that occurring during the sintering of ice
particles, since in both cases water molecules must be transported to reduce
the overall surface energy of the system. Studies of ice sintering have
shown that material transport via evaporation/deposition is substantially
more rapid than via surface or bulk diffusion processes (\cite{hobbs}, page
405), so we expect that evaporation/deposition will also be the dominant
transport mechanism in reaching the ice ECS. (In contrast, for many metals
that have been studied to date, surface diffusion is the dominant transport
process in the relaxation to the ECS. This is because these metals are often
chosen for their low vapor pressures, since this makes the experiments
simpler. The vapor pressure of lead near its melting point, for example, is
roughly $10^{10}$ times smaller than that of ice near its melting point.)

We can estimate the equilibration time via vapor transport by considering
first the case of a slightly prolate spheroid, with major axis length $%
2(R+\varepsilon )$ and minor axis length $2(R-\varepsilon )$. We further
assume an isotropic surface energy $\gamma $, so the ECS is a sphere. The
equilibrium vapor pressure just above any point on the surface of this
spheroid is%
\begin{equation}
c_{eq}\approx c_{sat}\left( 1+\frac{2\delta }{R_{eff}}\right)
\label{pressure}
\end{equation}%
where $c_{sat}$ is the saturated vapor pressure above a flat surface, $%
\delta =\gamma /c_{solid}kT\approx 1$ nm \cite{libbrechtreview} and $R_{eff}$
is the effective radius of curvature of the surface at that point. The vapor
pressure difference between the \textquotedblleft poles\textquotedblright\
and the \textquotedblleft equator\textquotedblright\ of the spheroid is then
approximately 
\[
\Delta c\approx \frac{2\delta \varepsilon }{R^{2}}c_{sat} 
\]%
and it is this difference that drives the system to its ECS. The growth rate
at the equator is then (using the notation in \cite{libbrechtreview})%
\[
v_{equator}\approx \alpha v_{kin}\frac{2\delta \varepsilon }{R^{2}} 
\]%
where $\alpha $ is the attachment coefficient at the growing surface. From
this we see that the change in shape of the spheroid is given approximately
by the differential equation%
\[
\frac{d\varepsilon }{dt}=-\alpha v_{kin}\frac{2\delta }{R^{2}}\varepsilon 
\]%
which has the solution $\varepsilon (t)=\varepsilon _{0}\exp (-t/\tau )$
with a time constant%
\begin{eqnarray}
\tau &=&\frac{R^{2}}{2\alpha v_{kin}\delta }  \label{tau} \\
&\approx &1.4\left( \frac{R}{100\textrm{ }\mu \textrm{m}}\right) ^{2}\left( 
\frac{1}{\alpha }\right) \left( \frac{1\textrm{ mm/sec}}{v_{kin}}\right) \textrm{
hours}  \nonumber
\end{eqnarray}

Since $v_{kin}\approx 1$ mm/sec near the melting point \cite{libbrechtreview}%
, we see that the equilibration times will be of order hours for a 100 $\mu $%
m crystal with $\alpha =1,$ and perhaps a week for a 1 mm crystal.
Unfortunately we frequently have $\alpha \ll 1,$ which then results in
longer equilibration times. We consider several special cases:

\textbf{Case 1:}\ If the crystal is equilibrating in a background of air or
some other inert gas, then $\alpha $ in Equation \ref{tau} should be
replaced with 
\[
\alpha _{eff}=\frac{\alpha \alpha _{diff}}{\alpha +\alpha _{diff}} 
\]%
where $\alpha _{eff}$ is an effective condensation coefficient that accounts
for the diffusion of water molecules through the air, and $\alpha _{diff}$
is described in \cite{libbrechtreview}. For a 100 $\mu $m crystal, $\alpha
_{diff}\approx 0.001$ with a background pressure of one atmosphere. Clearly
any ECS experiments with ice must be done in the absence or near absence of
any background gas, or the equilibration times will be exceedingly long.

\textbf{Case 2:}\ If our initial crystal is a completely faceted prism, then
reaching the ECS (assuming this is a nearly spherical shape) will require
removing water molecules from the corners of the prism and depositing them
on the facet surfaces. In this case there is a strong 2D nucleation barrier
that retards deposition on the facet surfaces, whereas there is no
nucleation barrier to removing molecules from the corners. Measurements of
the growth of facet surfaces reveal that typically $\alpha _{facet}\approx
\exp (-\sigma _{0}/\sigma ),$ where $\sigma $ is the supersaturation at the
surface and $\sigma _{0}$ is determined experimentally \cite{libbrecht}.
Using $\sigma \approx 2\delta \varepsilon /R^{2}$ from the above discussion,
we find prohibitively long equilibration times for crystals larger than even
a few $\mu $m.

\textbf{Case 3:}\ If we start with a perfectly spherical crystal, then
equilibrating to the ECS requires removing molecules from the facet surfaces
(as they become exposed en route to the ECS) and depositing them on the
rounded surfaces. In this case there is no nucleation barrier, and at low
background pressures we would have $\alpha \approx 1$ in Equation \ref{tau}.

Note that the nucleation barrier to reaching the ECS (Case 2) has also been
studied in metal systems \cite{mullins, rohrer, bonzel}. As with the ice
case, it has been found that a nucleation barrier can effectively prevent a
crystallite from reaching its ECS during the span of an experiment, and that
the equilibration time is a strong function of the initial crystal
morphology.

From these considerations we can identify a promising experimental path to
observing the ECS for ice. This begins with a small columnar crystal prism,
fully faceted, which can be easily grown. This crystal might be isolated by
suspending it in an electrodynamic trap \cite{swanson}, or perhaps by
placing the crystal on a suitable substrate in a sealed isothermal chamber.
A superhydrophobic surface, for example, might be expected to cause only
minimal perturbation of the equilibrium crystal shape.

The temperature of the experimental chamber would then be increased to near
the melting point, where the prism facets undergo a roughening transition 
\cite{elbaum}. This would remove the nucleation barrier on the prism facets,
and water molecules would thus be able to evaporate off the basal surfaces
and deposit onto the prism facets or onto rounded regions, all with $\alpha
\approx 1.$ During this process one could optically monitor the transition
to the ECS with time. Since the step energies are low near the ice melting
point, the ECS at high temperatures would be nearly spherical. Once the ECS
was obtained near the melting point, the temperature could then be lowered
to observe the ECS at lower temperatures (Case 3).

In addition to normal crystallites, it is also possible to observe the ECS
of \textquotedblleft negative\textquotedblright\ crystals, or voids in a
crystalline solid, as is commonly done in ECS studies. Here again we have
several illustrative cases:

\textbf{Case 4:}\ If the initial ice void is a faceted prism, as is the
typical shape when creating the void by evacuation \cite{furukawa}, then the
ECS must form by molecules being removed from the facet surfaces and
deposited in the corners. In this case the initial removal of molecules is
inhibited by a significant nucleation barrier, so the equilibration
timescale is prohibitively long.

\textbf{Case 5:}\ If the initial ice void is spherical in shape, then
molecules must be removed from the rounded surfaces and deposited on the
growing facets, and both processes will proceed with no nucleation barrier.

Here again we see a promising route to achieving the ECS in a negative
crystal. The process would begin with a columnar faceted void, which is a
known growth form during evacuation \cite{furukawa}. A temperature gradient
could be used to move the void away from its evacuation tube \cite{dadic},
thus isolating it inside the crystal in an evacuated state (i.e. with little
or no background gas present). At this point the temperature would be raised
above the prism roughening temperature, thus removing the nucleation
barrier, and again the transition to the ECS could be monitored. Once a
nearly spherical ECS was formed near the melting point, the temperature
could then be lowered as described in the positive crystal case.

We note in passing that this ECS experiment requires a high degree of
temperature uniformity. If an ice void of radius $R$ is exposed to a
temperature gradient $\nabla T$, then one side of the void will grow while
the other side evaporates, thus causing the void to move at a velocity%
\begin{eqnarray*}
v &\approx &\alpha v_{kin}\Delta \sigma \\
&\approx &\alpha v_{kin}\frac{R}{c_{sat}}\frac{dc_{sat}}{dT}\nabla T.
\end{eqnarray*}%
If we insist that a void moves by less than a distance $R$ over the
equilibration timescale, say 10 hours, then we must have $\nabla T<0.005$
C/cm, which is a nontrivial temperature gradient to achieve in an
experimental system.

We conclude that equilibration times of days or shorter could be achieved
for both normal crystallites and crystal voids, provided suitable
experimental conditions are applied. If a nucleation barrier inhibits the
transition to the ECS, however, then the equilibration times are
prohibitively long.

\section{Experiments to Date}

The ECS of ice has received little experimental attention thus far. Elbaum 
\cite{elbaum} reported observations of equilibrated ice crystals that were
largely spherical, with facets forming a small fraction of the total surface
area of the crystal. Out of equilibrium, the facets were found to be larger
during growth, and smaller during sublimation, as one would expect. Few
experimental details were given in this paper, unfortunately, so the reader
is left guessing about equilibration times and observations of the
transition from growth/melt forms to the ECS.

In contrast to the nearly spherical ECS observed by Elbaum, Colbeck reported
a fully faceted ECS for ice \cite{ice1}, but a careful reading of this paper
suggests several potential problems. For example, the crystals were fairly
large and grown in air, so one expects exceedingly long equilibration times.
Rounded plates were sometimes observed, and we would expect a substantial
nucleation barrier for such crystals to achieve a full equilibrium shape.
Also the experimental parameters were not especially well controlled, and
the transition to the ECS as a function of time was not monitored. It is
possible that these experiments were observing slow growth or evaporation
forms, and not the true ice equilibrium shapes.

In another experiment Furukawa and Kohata \cite{furukawa} observed the
formation of negative crystals, and these have been interpreted as showing a
fully faceted ECS when the evacuation process was very slow or halted \cite%
{garcke}. Again the paper does not elaborate on observations of the
equilibration times and the transition to the ECS, and again it appears
possible that growth forms were being observed.

Kobayashi's observations of ice crystal growth at low supersaturations \cite%
{kobayashi} are frequently cited in the context of the ice ECS or the
surface energy anisotropy, especially as a measurement of $\gamma
_{basal}/\gamma _{prism}$ \cite{pruppacher, hobbs, fletcher}. These
measurements were made in air using a supersaturation of typically a few
percent, however, which is substantially higher than the maximum
supersaturation for observing the ice ECS (see below). Thus again these
observations were likely of growth shapes, and not the true ice ECS.

In all these papers the lack of experimental details prevents one from
understanding their contradictory results. Better experiments are clearly
needed, and one would like to see careful observations of equilibration
times and the transition to the ECS, especially as a function of
temperature, crystal size, and initial conditions. Without this attention to
detail, and publications that report these experimental details, we are not
able to reach many definitive conclusions regarding the ice ECS.

The observations in \cite{ice2} also suggest the occurrence of pyramidal
facets on ice crystallites. However the images are somewhat unclear and the
equilibration times used were very short. Measurements of atmospheric halos,
and images of crystals collected during halo displays \cite{tape}, show that
the \{10$\overline{1}$1\} pyramidal facets do occur during ice crystal
growth at low temperatures, suggesting that these facets may also appear in
the ECS. However these observations were likely all of growth forms, so they
may in fact have no bearing on the ECS.

Numerous observations have been made of bubbles in ice, sometimes in the
laboratory and sometimes in very old ice formations \cite{carte, dadic,
chappellaz}. To my knowledge, however, neither spherical nor partially
faceted shapes have been seen in these samples, so it appears that these
bubbles also did not possess true equilibrium shapes.

My conclusion is that none of the observations or experiments to date has
conclusively demonstrated the ECS for ice. This is not surprising, since our
discussion above suggests that the equilibration times should be quite long
except in special circumstances. Plus there simply has not been much
experimental effort aimed at the ice ECS, in contrast to simple metal
systems. But although the ice ECS has not been definitively observed to
date, it does appear that successful experiments are possible if sufficient
care is taken.

\section{Ice Growth Dynamics}

Our final question, somewhat tangentially related to the ice ECS, is
regarding what role the anisotropic surface energy plays in ice crystal
growth from water vapor. We expect that if the growth rate is low enough, or
if the crystal size is small enough, then surface energy effects will begin
to overshadow kinetics effects in determining growth rates and morphologies.
Our question is then one of determining what physical mechanisms control the
ice growth behavior in what regions of parameter space.

Let us first consider the slow growth of a faceted ice crystal, a simple
hexagonal prism. We will assume the absence of any inert background gas, so
water vapor diffusion does not limit the growth. Since the equilibrium vapor
pressure of any point on the ice surface is given by Equation \ref{pressure}%
, we see that $c_{eq}$ will be highest at the corners of the prism. This is
a surface energy effect, and if the applied supersaturation $\sigma $ is
less than $\sigma _{corner}=\left( c_{eq,corner}-c_{sat}\right) /c_{sat},$
then the corners will become more rounded with time. However if $\sigma
>\sigma _{corner},$ then crystal growth will cause the corners to become
less rounded. Thus from Equation \ref{pressure} we see that the corner
radius for a growing crystal prism will be approximately%
\[
R_{corner}\approx \frac{2\delta }{\sigma }.
\]%
If $R_{corner}$ is greater than the overall size of the crystal, then the
crystal morphology will be essentially the ECS. At smaller values of $%
R_{corner},$ the morphology will be essentially an ice prism with slightly
rounded corners. Taking $R_{corner}=2$ $\mu $m in this expression gives $%
\sigma =0.1$ percent. Put another way, we expect that surface energy effects
will dominate only at supersaturations below a maximum of $\sigma _{\max
}\approx 2\delta /R_{eff},$ where $R_{eff}$ is the characteristic size of a
growing crystal feature. Typically $\sigma \gg \sigma _{\max }$ in ice
growth experiments, so we would expect surface energy effects are much
smaller than kinetic effects in determining the growth dynamics and crystal
morphologies.

If we add that the growth velocity is given by $v\approx \alpha
v_{kin}\sigma $ \cite{libbrechtreview}, the above expression tells us that
surface energy effects will begin to dominate if%
\begin{eqnarray*}
vR_{eff} &<&2\alpha \delta v_{kin} \\
\left( \frac{v}{1\textrm{ }\mu \textrm{m/sec}}\right) \left( \frac{R_{eff}}{1%
\textrm{ }\mu \textrm{m}}\right)  &<&\alpha 
\end{eqnarray*}%
where $R_{eff}$ is the size of a growth feature, and in the last expression
we took $v_{kin}=500$ $\mu $m/sec. Again, this inequality is not satisfied
in essentially all ice growth experiments done to date.

Another way of addressing this question is to consider the growth of a
dendrite tip, which has a growth velocity (see Equations 24-28 in \cite%
{libbrechtreview}) 
\begin{eqnarray}
v &\approx &\frac{2D}{RB}\frac{c_{sat}}{c_{solid}}\left[ \sigma _{\infty }-%
\frac{2\delta }{R}-\frac{v}{\alpha v_{kin}}\right]  \label{vel} \\
&\approx &\frac{2D}{RB}\frac{c_{sat}}{c_{solid}}\left[ \sigma _{\infty }-%
\frac{2\delta }{R}-\frac{\sigma _{\infty }R_{kin}}{\alpha R}\right] 
\nonumber
\end{eqnarray}%
where%
\[
R_{kin}=\frac{2D}{B}\sqrt{\frac{2\pi m}{kT}}\approx 30\textrm{ nm} 
\]%
For typical atmospheric growth we have $\sigma _{\infty }/\alpha \left(
\sigma _{surf}\right) >1,$ so the kinetics term in Equation \ref{vel} is
roughly an order of magnitude larger than the surface energy term. Using
direct measurements of dendritic growth (see the discussion following
Equation 31 in \cite{libbrechtreview}) again shows that the kinetics term is
roughly an order of magnitude larger then the surface energy term.

Our overall conclusion from this analysis is that attachment kinetics are
more important than surface energy in determining ice crystal growth rates
from water vapor, by roughly an order of magnitude for typical experimental
or atmospheric conditions. Furthermore, to my knowledge no experiments to
date investigating ice crystal growth from water vapor have conclusively
observed any surface energy effects. In part this is because vapor diffusion
and surface attachment kinetics are dominant effects, but also evaporation
(with $\sigma <0)$ can produce rounded shapes that are essentially
indistinguishable from surface energy rounding (which occurs during growth,
when $\sigma $ is small but still positive). As with the ECS, measurements
of surface energy effects in ice growth dynamics from the vapor phase are
likely observable in conditions of very slow growth of very small crystals,
but so far have not been definitively observed.

\section{Conclusions}

Based on the above analysis, we reach the following conclusions regarding
the ice ECS, the anisotropic surface energy, and surface energy effects on
ice crystal growth from water vapor.

\textbf{The ECS has not yet been definitively observed for ice crystals.}
This conclusion stems from the many observations of the ECS in metal
systems, our theoretical understanding of those systems and its application
to the case of ice in water vapor, and the fact that the experiments to date
have not made a convincing case for the observation of the ice ECS. An
examination of ECS measurements in metals suggests that the experiments are
quite difficult, and it is easy to mistake growth/evaporation forms for the
ECS. Experiments to date in ice have not monitored the transition to the ECS
as a function of time, and have probably underestimated the time needed to
reach the ECS.

\textbf{The surface energy anisotropy is likely cusp-like near the faceted
surfaces, and the size of the cusps can be estimated from crystal growth
measurements.} Again this conclusion stems in part from our theoretical
understanding of the ECS in other systems, which in turn comes from ECS
experiments. Comparing with metal systems and the step energy measurements
described above, the evidence suggests that $(\gamma _{\max }-\gamma _{\min
})/\gamma _{ave}$ in ice is no larger than 5-10 percent. The standard model
for the ECS (developed largely from metal systems) includes cusps in $\gamma
(\hat{n})$ at the facet angles. It is highly likely that this model applies
to ice as well, so that the sizes of the facet cusps can be estimated from
step energy measurements.

\textbf{The equilibrium shape is likely nearly spherical with only small
faceted regions.} Measurements of ice step energies allow an estimate of the
cusps in $\gamma (\hat{n})$, and the sizes of the faceted regions follow
from the Wulff construction, as described above. The result is that only a
small fraction of the ice ECS is faceted, with the fraction becoming larger
at lower temperatures.

Given the poor state of ice ECS experiments, it remains possible (I believe
somewhat remotely) that $\gamma _{basal}$ and $\gamma _{prism}$ are both
substantially lower than $\gamma (\hat{n})$ at angles far from the facet
angles, and in this case the ECS would indeed be fully faceted, or nearly
so. However the preponderance of the evidence suggests that the surface
energy anisotropy is small. The fact that ice does not readily cleave along
facet surfaces further supports this conclusion. Accurate observations of
the\ ECS could remove the remaining uncertainties.

\textbf{The time needed to reach equilibrium is likely prohibitively long,
except under special circumstances}. As described above, there are
substantial 2D nucleation barriers that prevent the growth of facet surfaces
at low supersaturations. These nucleation barriers have been observed in ice
growth measurements, which reveal their size and temperature dependence.
(Note that this assumes the absence of any dislocations, which was one of
our initial assumptions in this examination of the ice ECS. This assumption
is warranted from an experimental standpoint because growth measurements
have shown that dislocations are relatively rare in small ice crystals \cite%
{libbrecht}.) The same nucleation barriers are present in metal ECS systems,
and they have been shown to have substantial effects on the equilibrium
times in these systems. However, the roughening transition on the prism
facets for $T>-2$ C gives a possible route to achieving the ECS in ice, as
described above.

\textbf{Surface energy effects likely play a relatively smaller role in ice
crystal growth dynamics when compared to the role of attachment kinetics}.
This conclusion follows from an examination of ice crystal growth
experiments and ice growth under atmospheric conditions, as described above.
Effects from attachment kinetics are roughly an order of magnitude larger
than effects from surface energy under typical conditions. Surface energy
effects are expected to dominate only for extremely small crystals (of order
microns in size or smaller) and when the growth is extremely slow (so that
the growth morphology is only slightly changed from the ECS). In addition,
the surface energy anisotropy is likely small, of order a few percent, while
the anisotropy in the attachment kinetics is typically far greater. Since
many morphological features arise from anisotropies, this further supports
the statement that attachment kinetics is substantially more important than
surface energy effects in governing ice growth.

\end{document}